\documentclass[10pt,twocolumn,letterpaper]{article}

\usepackage{ijcb}
\usepackage{times}
\usepackage{epsfig}
\usepackage{graphicx}
\usepackage{amsmath}
\usepackage{amssymb}
\usepackage[norule,symbol,perpage]{footmisc}

\usepackage{comment}
\usepackage{color}
\usepackage{url}
\usepackage{subfigure}
\usepackage{gensymb}
\usepackage{siunitx}
\usepackage{cite}
\usepackage{float}

\ijcbfinalcopy 

\ifijcbfinal\fi

\makeatletter
\def\ps@IEEEtitlepagestyle{
\def\@oddfoot{\mycopyrightnotice}
\def\@evenfoot{}
}
\def\mycopyrightnotice{
{\hfill \footnotesize 978-1-7281-9186-7/20/\$31.00 \copyright 2020 IEEE\hfill}
}
\makeatother

\begin{document}

\title{All-in-Focus Iris Camera With a Great Capture Volume}

\author{Kunbo Zhang\textsuperscript{\textit{a}}, Zhenteng Shen\textsuperscript{\textit{b}}, Yunlong Wang\textsuperscript{\textit{a}} and Zhenan Sun\textsuperscript{\textit{a}}\\
\textsuperscript{\textit{a}} Center for Research on Intelligent Perception and Computing\\
National Lab of Pattern Recognition, Institute of Automation\\
Chinese Academy of Sciences, Beijing, China  \\
\textsuperscript{\textit{b}} Tianjin Academy for Intelligent Recognition Technologies, Tianjin, China\\
{\tt\small kunbo.zhang@ia.ac.cn, shenzhenteng@tj.ia.ac.cn, yunlong.wang@cripac.ia.ac.cn,}\\
{\tt\small znsun@nlpr.ia.ac.cn}

}

\maketitle
\thispagestyle{empty}

\begin{abstract}
   Imaging volume of an iris recognition system has been restricting the throughput and cooperation convenience in biometric applications. Numerous improvement trials are still impractical to supersede the dominant fixed-focus lens in stand-off iris recognition due to incremental performance increase and complicated optical design. In this study, we develop a novel all-in-focus iris imaging system using a focus-tunable lens and a 2D steering mirror to greatly extend capture volume by spatiotemporal multiplexing method. Our iris imaging depth of field extension system requires no mechanical motion and is capable to adjust the focal plane at extremely high speed. In addition, the motorized reflection mirror adaptively steers the light beam to extend the horizontal and vertical field of views in an active manner. The proposed all-in-focus iris camera increases the depth of field up to 3.9 m which is a factor of 37.5 compared with conventional long focal lens. We also experimentally demonstrate the capability of this 3D light beam steering imaging system in real-time multi-person iris refocusing using dynamic focal stacks and the potential of continuous iris recognition for moving participants.
\end{abstract}

\let\thefootnote\relax\footnotetext{\mycopyrightnotice}

\section{Introduction}

\begin{figure}[!t]
\centering
\includegraphics[scale=0.25]{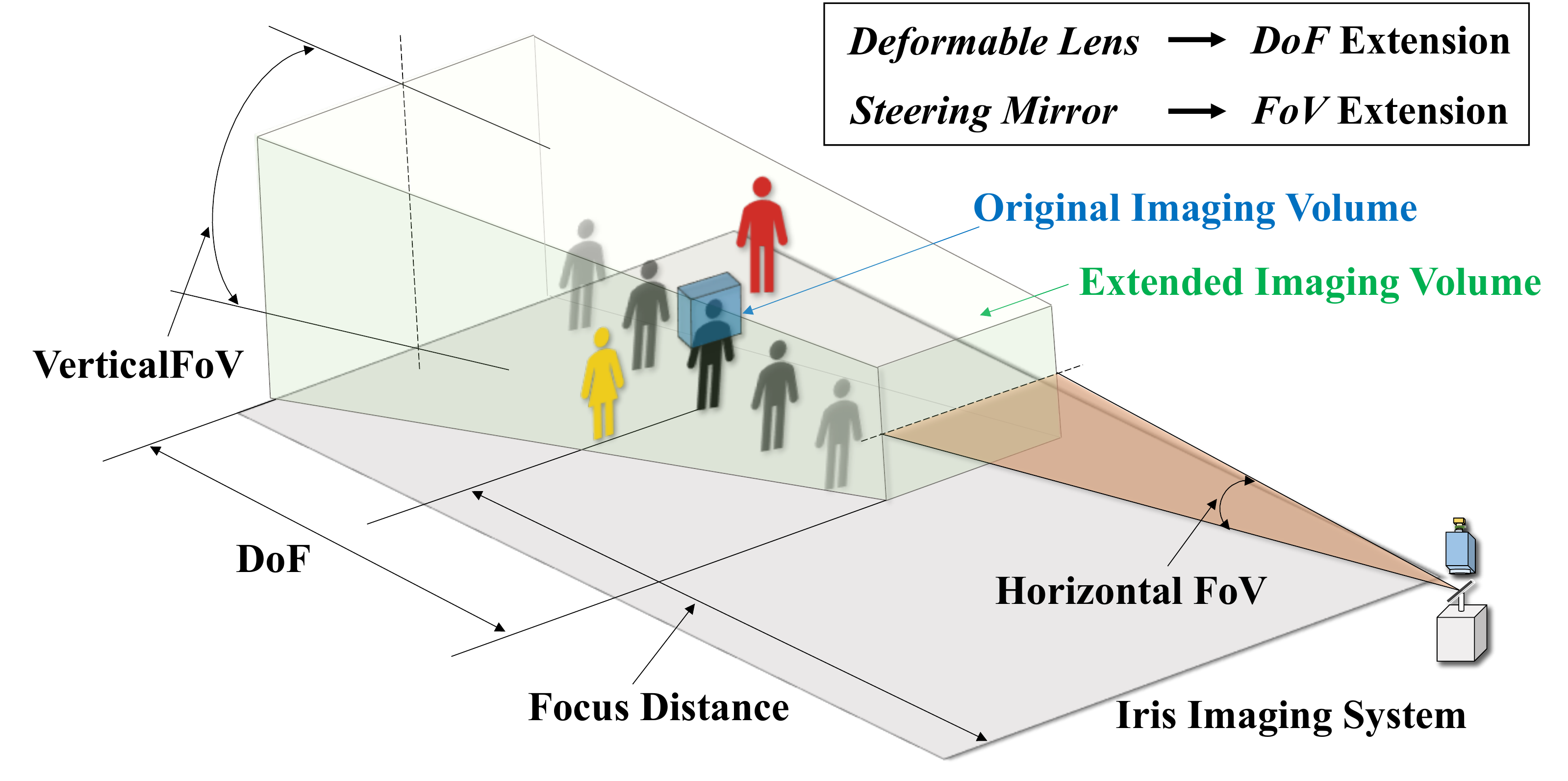}
\caption{\textbf{Capture volume extension in an iris imaging system.} DoF extension allows a participant to stand at various distances. FoV expansion extends the capture capability of participants at various orientations with different heights.}
\label{fig_ImagingVolume}
\end{figure}

The capture volume of an imaging system refers to the size of space in which the scene points appear in acceptably sharp focus in visual information acquisition. Large capture volume is a longstanding goal in many optical imaging related applications such as tomography, biometrics, and microscopy. Limited capture volume has been the bottleneck of iris recognition throughput improvement compared with the rapid development in processing algorithm. Increasing the range of depth of field (DoF) and field of view (FoV) will allow less cooperation from participants \cite{Ross_LongRangeSurvey}. To extend DoF, one can reduce the aperture size at the expense of decrease in light collection and signal-to-noise ratio (SNR) \cite{Ma_Beyond}. Other capture volume extension trials have focused on multiple cameras \cite{Matey_IOM}, camera rotation \cite{PTZCVPR}, and zoom lens \cite{Ross_LongRangeSurvey} in iris imaging. Since there is a trade-off between  resolution and capture volume as shown in Figure \ref{fig_ImagingVolume}, the dominant optical design is still fixed-focus camera in industry. Focal Sweep and multiplexing have been proposed as techniques to extend the capture volume of an imaging system while maintaining high resolution and fast response \cite{Nayar_FocalSweep, Helmes_TemporalMultiplexing}. 

Mechanical motion based focal sweep was proposed by Zhou \cite{Zhou_MotionSweep} using a micro-actuator for seamless refocusing. Following this idea, recent advances have been able to extend the depth range on the order of 100's of centimeter. Telescope imaging was introduced to achieve larger standoff distance on the order of 1's of meter in range distance iris recognition \cite{Ve_LongFocal}. Bulky optical and focus control components, however, implies critical application issues. Telephoto zoom iris imaging is slow due to mechanical motion and long travel distance with uncertain lifetime\cite{Gao_ZoomLens}. In this paper, we seek to overcome such drawbacks by integrating a deformable liquid lens with existing system to demonstrate the advantages of focal sweep in DoF extension of iris recognition. Such a bio-inspired imaging mechanism is similar to human vision which changes the focal point using ciliary muscle to adjust lens curvature instead of moving the position of lens \cite{Nilson_HumanEye}.

FoV of a singlet or zoom lens is fixed and narrow in telephoto optical system. To extend the field of view, Funahasahi  \cite{Ko_PTZ} developed a pan-tilt-zoom (PTZ) iris imaging device to track human head motion in an expanded area. As object distance increases, the weight and travel distance of lens components slow the time response in dynamic scene. Digital micromirror device as a light and compact alternative has been widely used in micro-scale imaging \cite{To_MEMSEndscope} to observe wider field while maintaining high resolution. Even off-the-shelf MEMS mirror size is not sufficient for iris imaging, it is proven to capture temporal multiplexing parallel videos of four different objects using a single camera for almost real-time \cite{Helmes_TemporalMultiplexing}. We focus our research on steering mirror based FoV extension together with focus-tunable lens based DoF extension to expand iris capture volume over all three dimensions. 

The following are the major contributions of this paper.

\begin{figure}[!t]
\centering
\includegraphics[scale=0.5]{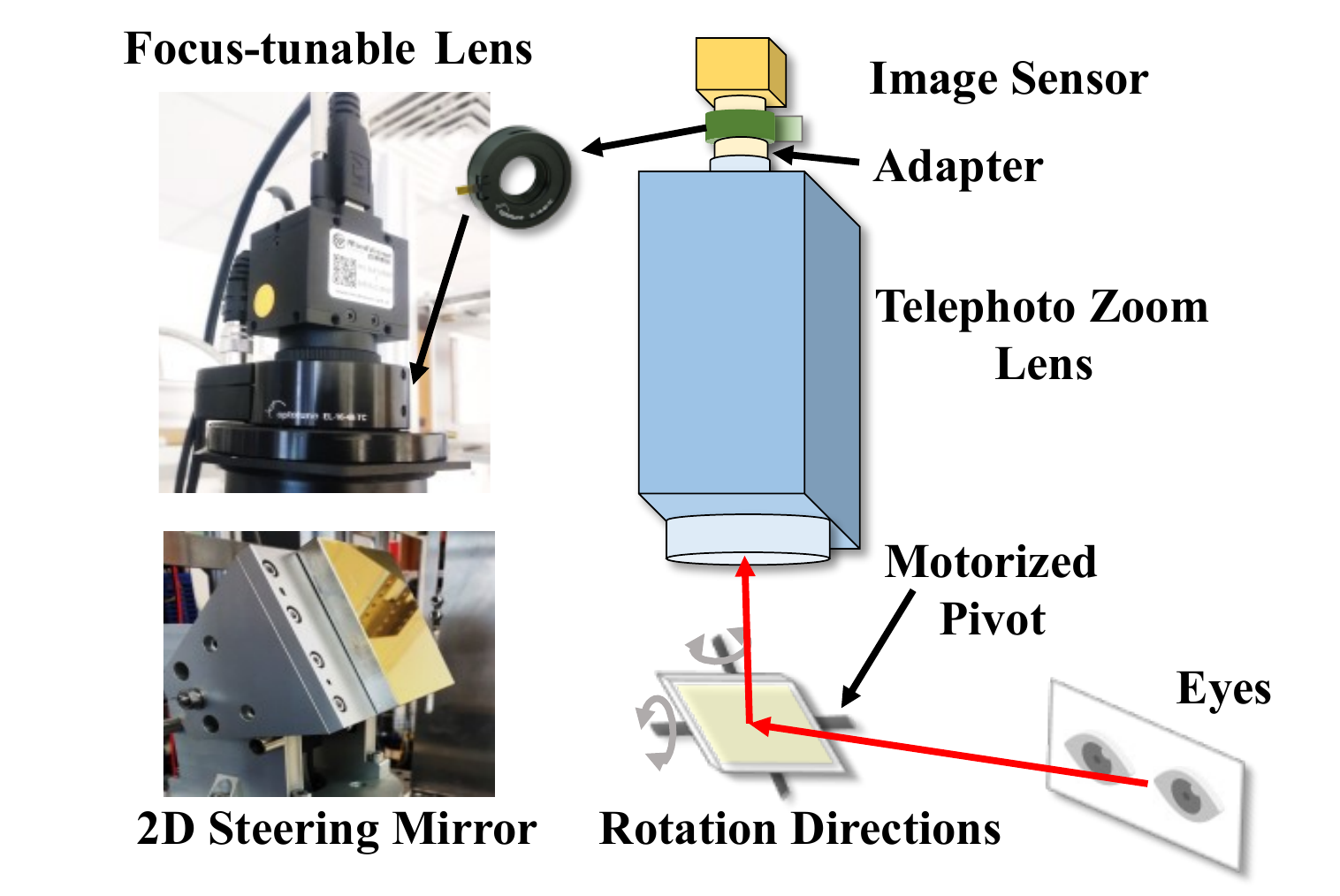}
\caption{\textbf{Overview of our all-in-focus iris imaging system schematic.} The telephoto zoom lens and image sensor are placed vertically downwards. The focus-tunable lens is mounted between the zoom lens and the image sensor with a customized ring adapter to compensate back focal length. The reflection mirror is motorized to steer the light back from the eyes into the sensor by two-dimensional rotation.}
\label{fig_system}
\end{figure}

\textbf{Focus-tunable lens for DoV extension.} We present a focus-tunable iris camera by integrating a deformable lens with a telephoto zoom lens. This new imaging device is able to adjust the desired focal plane by electrically changing optical power with superior dynamic response as shown in Figure \ref{fig_system}. A comparable variable focal length iris imaging solution requires complicate design of lens groups and mechanical motion which introduce significant time delay and challenges in durability.

\textbf{Adaptive 2D steering mirror for FoV extension.} The introduction of an adjustable mirror allows the precise light path manipulation with fast response for temporal multiplexing iris imaging. Instead of rotating the heavy telephoto lens, the reflection mirror with gold-coated fine processed surface is agile in motion control. Literally this 2D steering mirror can capture a participant at any position in a 360\degree surrounding.

\textbf{All-in-focus iris camera with a great capture volume.} Previous long range iris imaging researches are mostly restricted by its extension capability. In this study, we demonstrate that our compact deformable lens increases the DoF 37.5 times and the steering mirror expands the FoV to $\pm$180\degree (H) and $\pm$60\degree (V) for stand-off iris recognition. Our proposed all-in-focus iris imaging system supersedes traditional capture volume extension methods such as PTZ and wavefront coding in deployment and user cooperation. 

\textbf{Multi-person iris refocusing.} The great capture volume of our proposed system extends application scenarios of iris recognition. We demonstrate the usage of spatiotemporal multiplexing in iris imaging by examples such as real time multi-person auto-refocusing and continuous iris capture for a moving object.

\section{RELATED WORK}

As one of the critical measures of iris recognition system performance, capture volume is the 3D volume in which the imaging device capable to capture sufficiently qualified eye images of a participant \cite{Ross_LongRangeSurvey}. Larger capture volume in free space enables less user cooperation and better recognition performance. The longitudinal and the transverse dimensions in capture volume are defined by DoF and FoV respectively. Most capture volume extension works concentrate on extending DoF by engineering innovative optical elements such as wavefront coding, focal sweep, telescope zoom lens, and light field camera \cite{Pl_CompIris, Zhang_LightIris}. Traditional DoF extension approach is to increase F-number by using longer focal length or smaller aperture. 

The first known multi-biometric acquisition system with a 3-6 meters stand-off range uses the long focal zoom lens design to extend the focus point by mechanical adjustment of group zoom lens as illustrated in \cite{Bashir_EagleEye}. Considering the bulky size and slow reaction, new computational imaging method is introduced for rapid biometric information acquisition \cite{Zhang_LightIris}. A light-field camera uses computational based auto-refocusing method for DoF extension to overcome the trading-off between DoF and aperture size in a conventional camera. However, the micro-lens based light field imaging comes at the cost of a sacrifice in spatial resolution and difficulties in fabrication. Recent works by Dong \emph{et al.} \cite{Dong_Wavefront} and Hsieh \emph{et al.} \cite{Hsieh_Wavefront} use wavefront coding and post-processing to compensate the blurring caused by phase mask for DoF extension. However, necessary image restoration algorithm and cubic phase mask optimization increase the complexity of system implementation to maintain invariant point spread function (PSF) \cite{Smith_Wavefront}. The limited DoF extension capability (6cm to 18cm) makes such kind of computational imaging techniques impractical in complex scenes \cite{Pl_CompIris}. 

Since longer focal lens covers a smaller view angle, methods such as pan-tilt-zoom (PTZ) and multiple cameras are proposed to expand the FoV in iris imaging \cite{Wilburn_CamArray}. Venugopalan \cite{Ve_LongFocal} uses an off-the-shelf PTZ camera with the 360\degree \ pan and 180\degree \ tilt capabilities to capture unconstrained iris image between 0.5 m and 1.6 m. Even the pan-tilt movement with an adjustable speed up to 450\degree /second, the heavy and bulky optical form factors will jeopardize the performance of response time and angular resolution when it comes to distant object. Another intuitive solution to expand the view angle of iris image capture is by increasing the number of cameras as demonstrated in the well-known IOM system from Sarnoff in 2006 \cite{Matey_IOM}. In this IOM system the 20 cm capture height of a single camera is increased to 37 cm and 70 cm by vertically stacking two cameras and four cameras separately. If a camera matrix is used to cover large view angle, the geometry size and calibration work will become an obstacle that cannot be ignored \cite{Bok_CamArrayCali}.        

In contrast, our all-in-focus iris capture system can extend DoF with much larger ranges (\SI{3.9}{\metre} when focal plane is set at 5m) and 360\degree \ omni-directional view without participant height restriction by manipulating light beam direction in three-dimensional space. To extend DoF, a focus-tunable liquid lens is used in this iris camera by sweeping focal planes at extremely high speed without any mechanical motion and optical system modification. Miau \cite{Nayar_FocalSweep} extends the DoF of face video more than 10\emph{m} at 20\emph{fps} using periodic focal stacks acquired through deformable lens. To extend the horizontal and vertical FoVs, we steer a fast 2D reflection mirror instead of the heavy telephoto components. Optical multiplexing imaging has been successfully demonstrated to expand FoV in microscopy imaging by controlling steering mirror to superimpose images from multiple FoVs onto a single focal plane \cite{Stir_BrainMulti}. The major advantage of this system design is the dramatically increased capture volume while maintaining high spatial, temporal and angular resolutions for distant iris imaging.

\section{PROPOSED IRIS IMAGING SYSTEM}

To design the hardware for a distant iris imaging system, several fundamental components such as imaging sensor, lens, and illumination should be taken into consideration. In this section, the hardware configurations are discussed in terms of evaluating the capability to extend iris capture volume towards better iris image quality acquisition. The average diameter of human irides is on the order of 1\,$cm$. Even the National Institute of Science and Technology (NIST) showed that 120\,$pixels$ across iris diameter is suitable for recognition \cite{NIST7820}, we will still employ the 200\,$pixels$ across the iris region from ISO standard as the threshold in this study according to our research experience on long distance scenario \cite{ISO19794-6} . We explain the design decisions leading us to build the major components in this prototype imaging system. And we conclude this section by describing how large the capture volume is expanded in 3D by combining the focus-tunable lens and 2D steering mirror in theoretical calculation.

In this paper, our prototye system uses an image sensor with pixel numbers $4080\times3072$ at 30.5\,$fps$ and 35\,$\%$ relative response at 850\,$nm$. In order to investigate the capture volume change, we designed and manufactured an infrared zoom lens to cover the distance between 1\,$m$ and 5\,$m$ with a variable focal length between 70\,$mm$ and 350\,$mm$. Such a 15-piece \emph{f\,/\,4.8} lens has view angles of 18\degree x18\degree \ at 70\,$mm$ focal length and 3.8\degree x3.8\degree at 350\,$mm$. According to the lens equation under paraxial approximation, the DoF of an imaging system can be estimated as:

\begin{equation}\label{eq1}
    DoF = D_N + D_F = \frac{2Cd}{fP/(d-f)-C^2(d-f)/fP}
\end{equation}

\noindent where the near limit $D_N$ and far limit $D_F$ are a function of circle of confusion $C$, subject to lens distance $d$, focal length $f$, and exit pupil $P$. When the focal plane is set at 5\,$m$, the DoF  can be derived as 91\,$mm$ where $f$ = 350\,$mm$, $d$ = 5000\,$mm$ and the F-number of the lens is 4.8. Hence the capture volume is defined by the size of $DoF$ x $FoV$ which is approximately 0.04\,$m^3$

Since the intricate iris texture is well captured at wavelength between 700 and 900 \emph{mm} for Asian eyes, we have designed an adaptive illuminator to provide uniform coverage over the entire frontal view between 1 \emph{m} and 5 \emph{m} for any participant not higher than 2.2 \emph{m}. This 850 \emph{mm} illuminator is designed as matrix of individually controlled illumination modules which meet the Class I LED safety requirements in IEC standard 60825-1. When the participant eye is detected in the permissible space, only the illumination module governing the specified volume is turned on with a duration time matching the sensor exposure time.

\begin{figure}[!t]
\centering
\includegraphics[scale=0.32]{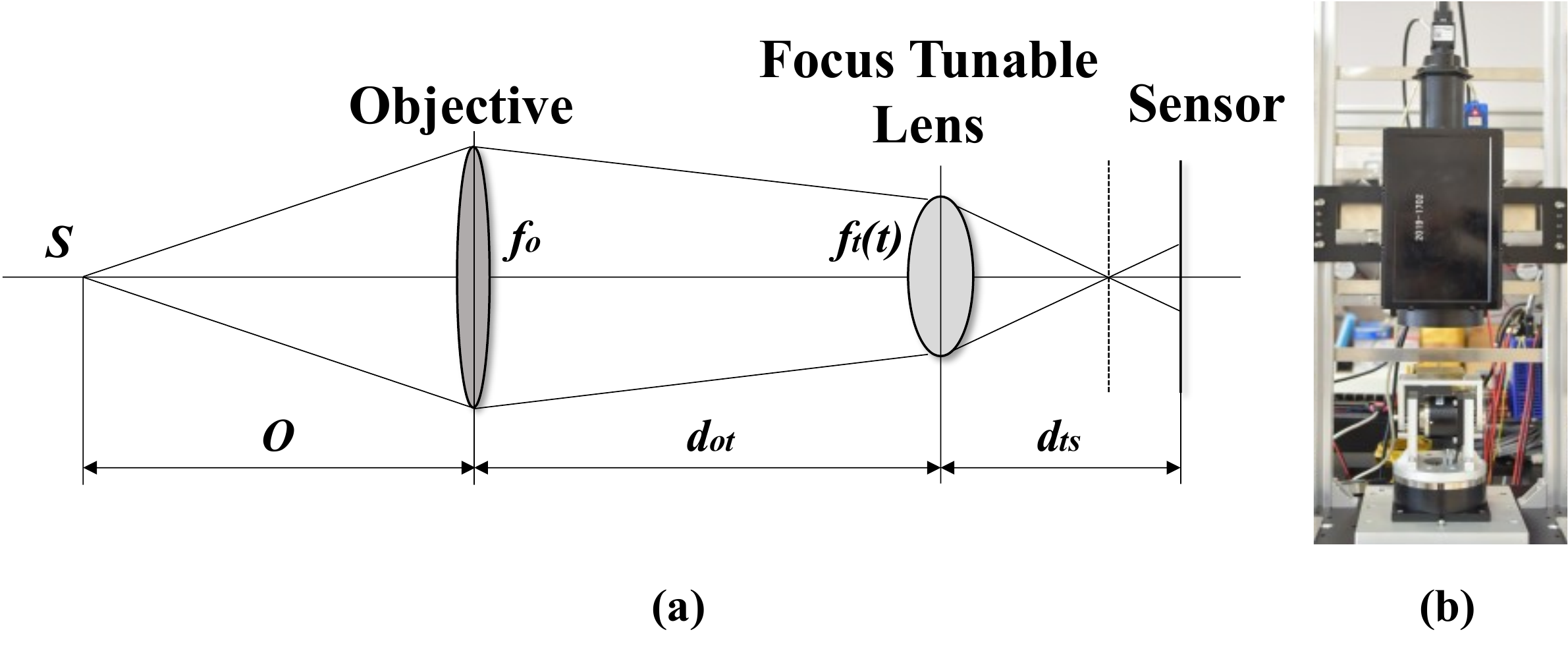}
\caption{\textbf{The focus-tunable lens model (a) and the 3D light beam steering prototype system (b).} We use a compound system of two thin lens to model our imaging system. The main optical axis of the lens is aligned with the center of the reflection mirror.}
\label{fig_lensmodel}
\end{figure}

\subsection{Focus-Tunable Lens for DoF Extension}

Compared with a conventional singlet lens with fixed focal distance, the focal length of a focus-tunable lens is adjustable dynamically. In our imaging system, we focus on deformable surface approach due to its large aperture size and superior dynamic response in which the membrane shape is changed by applying an electric current. We choose the Optotune tunable lens EL-16-TC-20D with a clear aperture of 16 \emph{mm}. The major hallmark of its large aperture benefits the collection of infrared light reflected back from the eyes of a participant at a long standoff distance \cite{Optotune}. 

The optical power of this focus-tunable lens is controlled by changing the electrical current flowing through the coil of the actuator over an optical power range of -10 \emph{dpt} and +10 \emph{dpt} corresponding to -100 \emph{mm} and 100 \emph{mm} within conserved temperature region. Both the optical fluid and the membrane material are highly transparent in the range of 400 and 2500 \emph{nm} which is ideal for iris imaging. The lens control driver is able to set the current with a 0.07\,$mA$ step  with frequencies from 0.2 to 2000 Hz. The response and settling times are 5 \emph{ms} and 25 \emph{ms} which are magnitude faster than most mechanical alternatives. We manufactured an additional spacer plus the general c-mount to attach this tunable lens between our long focal lens and the image sensor with optimal focusing result on imaging plane. In practice, we typically use 1\,$mA$ as the minimal step size for focus adjustment which roughly equal to 1 \emph{cm} in the depth direction. Since the liquid lens repeatability error of $\pm$0.1 dpt does not have obvious effect on the quality of eye image for recognition, we did not employ any control strategy for compensation.

To estimate the focal length of our optical components, we model our implementation as a compound system of two thin lens with one objective and one focus-tunable lens. Such an approximation is reasonable because most geometric aberrations have been corrected in our designed zoom lens which is simple for modelling. It is required to apply a large separation between the principal planes of the focus-tunable lens and the telephoto lens to minimize geometric aberrations. The focal length for the combined system is given by:

\begin{equation}\label{eq2}
    \frac{1}{f} = \frac{1}{f_o} + \frac{1}{f_t} - \frac{d_{ot}}{f_of_t}
\end{equation}

\noindent where $f$ is the focal length of the combined system, $f_o$ is the focal length of the zoom lens, $f_t$ is the focal length of the focus-tunable lens and $d_{ot}$ is the distance between the principal planes of the zoom lens and the focus-tunable lens. The value of $f_t(t)$ is a time-varying function as control current changes. When the object distance is 5000\,$mm$, the extended back focal length is derived as 247\,$mm$ considering $f_o$ = 350\,$mm$. When the object distance is much larger than objective lens, the focal length of the tunable lens $f_t(t)$ is independent of object distance $o$.

\begin{figure}[!t]
\centering
\includegraphics[scale=0.4]{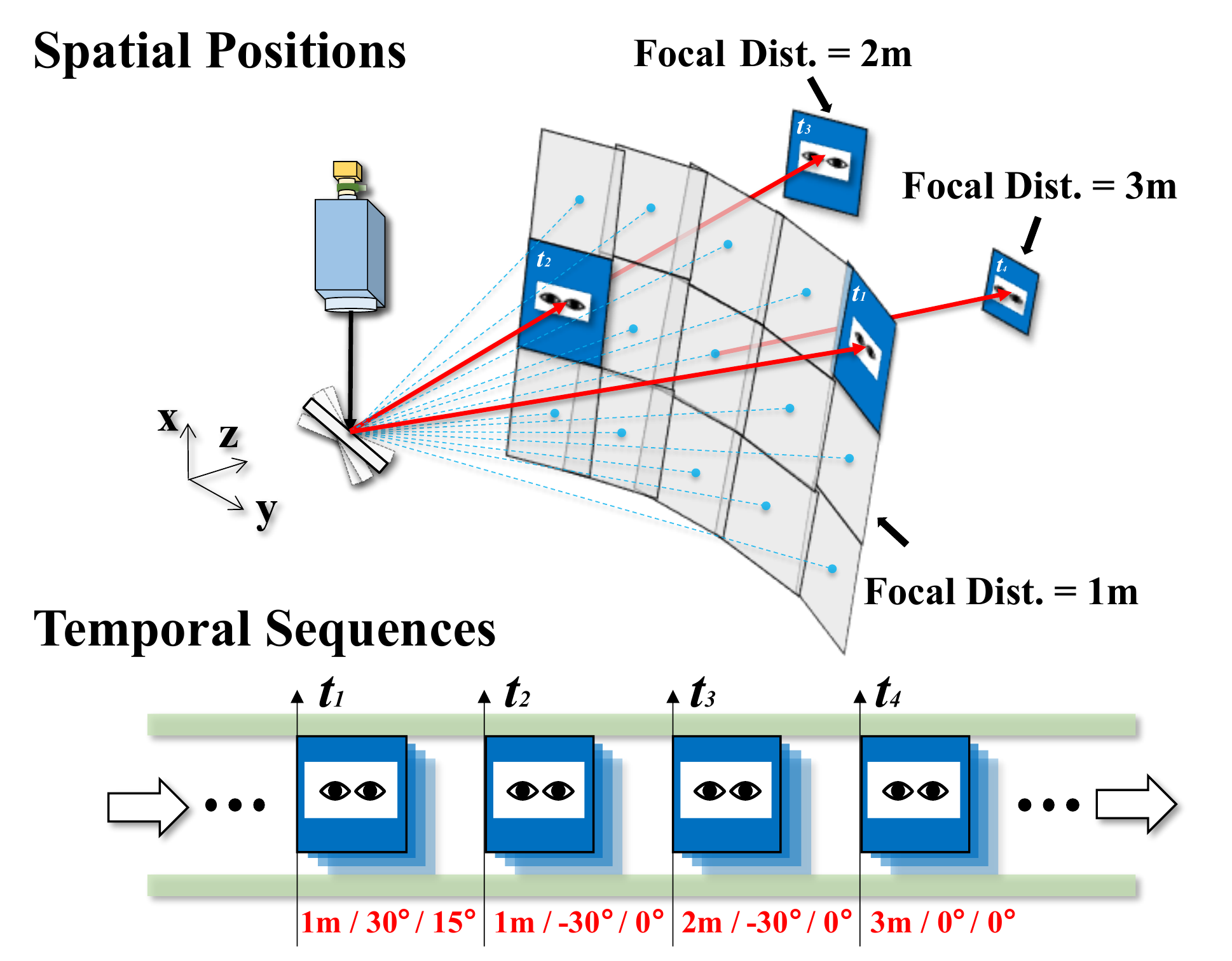}
\caption{\textbf{Spatial and temporal multiplexing imaging for iris capture.} Four pair of eyes at different spatial positions are sequentially captured by joint control of the liquid lens and steering mirror. From the capture starting time $t_i$, it may take several trials to acquire one qualified image before refocusing to the next subject. The first and second pairs of eyes are both located at 1\,$m$ distance with different view angles. All four pairs of eyes are captured using the same image sensor in a controlled sequence.}
\label{fig_multiplexing}
\end{figure}

\begin{figure}[!t]
\centering
\includegraphics[scale=0.3]{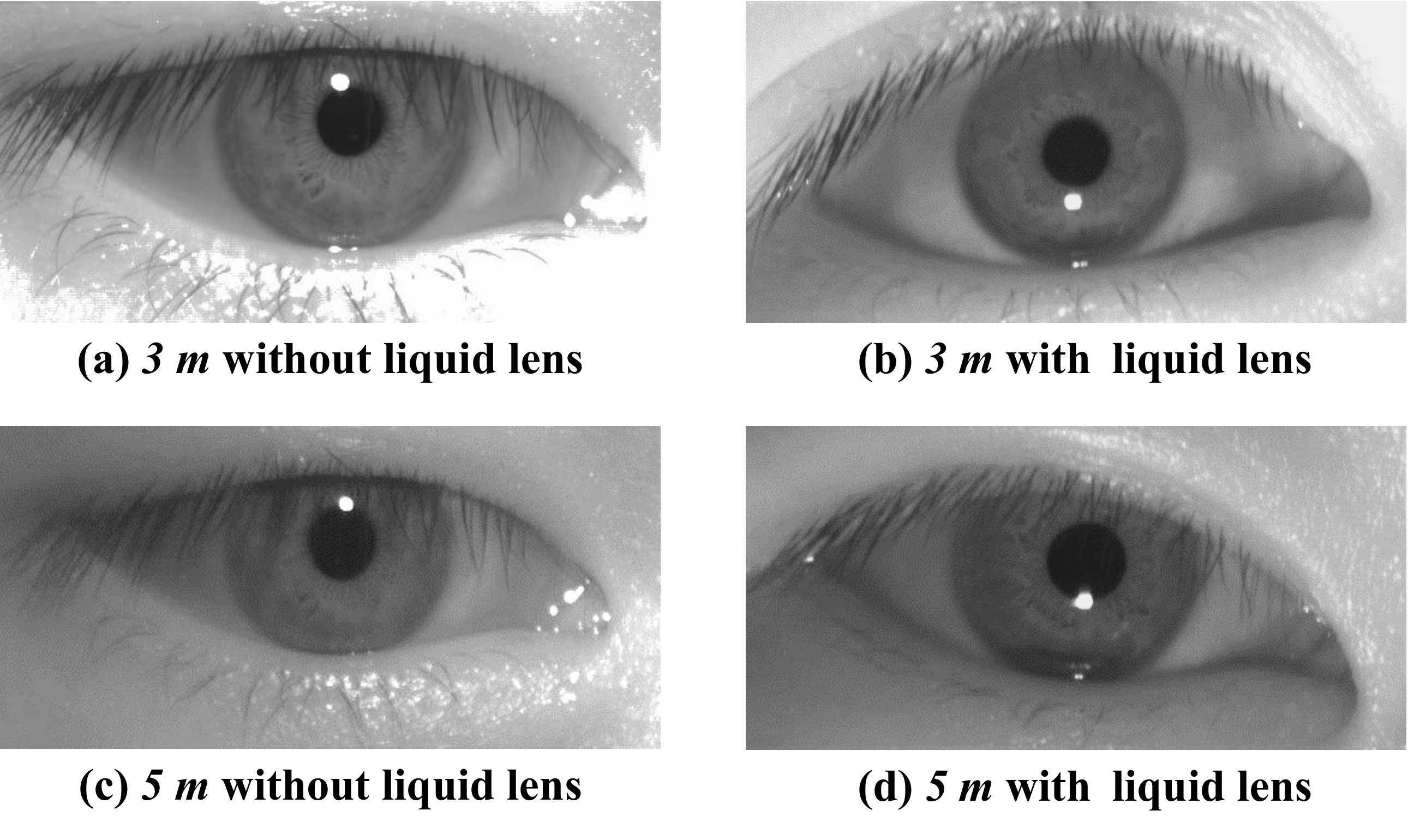}
\caption{\textbf{Comparison of iris image quality captured at two distances to evaluate the effect of focus-tunable lens.} The iris images (b) and (d) acquired using liquid lens appear darker due to infrared light transmission loss.}
\label{fig_ResolutionComp}
\end{figure}

To evaluate the impact of the liquid lens on image resolution, we have compared the iris image with and without it at 3\,$m$ and 5\,$m$ two distances as shown in Figure \ref{fig_ResolutionComp}. Although less photons are collected after the beam passes through the extra focus-tunable lens resulting in a relatively dark iris image. It is demonstrated that the details of the iris structure are preserved regardless of standoff distance and curvature of the liquid lens membrane. Even the focal stacks are not simultaneous captured, it is feasible to use this focus-tunable lens to capture multiples eyes in real-time while maintaining the high resolution quality of each image due to its fast response.

\subsection{2D Steering Mirror for FoV Extension}

Besides the fast focus-tunable method for DoF extension along depth direction, it is also important to expand horizontal and vertical views for iris imaging of unrestricted standing position. Based on the fundamental reflection law, we have introduced a 2D light beam steering method for FoV expansion by multiplexing imaging into one sensor in a temporal sequence. Such a fixed inverted lens architecture has been widely used in microscopy imaging and laser beam control. Compared with the traditional PTZ method to rotate the heavy zoom lens, the 2D motorized high reflective mirror offers precise and fast manipulation of light ray direction. Due to the small diameter, the common MEMS scanner and galvo mirror are excluded even they are fast up to several kilohertz. Considering our telephoto lens with an entrance pupil up to 73\,$mm$ for reflection light collection, we have designed a large size mirror driven by two high speed motors with exceptionally large tilt angle.

The new 2D steering device will steer the reflected light from human iris in the addressed imaging volume back into the sensor chip on focus. This customized 2D steering mirror can scan large 2D angles which is literally $\pm$180\degree \ in horizontal and over $\pm$60\degree \ in vertical. Compared with the small 3.8\degree x3.8\degree \ view angle of telephoto zoom lens at 5\,$m$, this new design is able to capture a participant standing at any position as long as the infrared lighting covers. To maximize the entrancing light, we have designed the reflection mirror as 70\,$mm$x60\,$mm$x10\,$mm$ size with high reflectance ($\geq$ 95\,$\%$) gold coating for 850\,$nm$ infrared light. The angular resolution of horizontal and vertical is 0.01\degree \ with a rotational speed up to 3500 rpm. The control signal pulse frequency is up to 1\,$MHz$ which delivers fast response capability.

Combining a 2D steering mirror with an electrically focus-tunable lens, our all-in-focus capture volume extended iris imaging system enables precise and fast light beam collection from any area within the addressable 3D space. Taking advantage of the extremely fast system response, the participants at various spatial positions are sequentially captured to form up the real-time frame sequence as shown in Figure \ref{fig_multiplexing}. It is noted that the four participants located at different distance (1-3\,$m$) and different angles (-30\degree \ to 30\degree). By jointly adjusting the focus-tunable lens and 2D steer device, our imaging system sequentially capture all  four eyes from $t_1$ to $t_4$. From the imaging staring time of one participant $t_i$, it may take several frames to capture a qualified image before adjusting the device for next participant because of user cooperation and scene change. The targeted iris image will come into focus at least one frame of the temporal focal stacks. The spatial resolution of our volume extension device is majorly determined by the step resolution in focus-tunable lens and angular resolution of the 2D steering mirror. The temporal resolution is determined by the time response of the focus-tunable lens and the steering mirror as well as the image sensor exposure time. When it comes to a moving object, this imaging system will become effective only if equipped with fast scene understanding method and precise control algorithm.  

\section{EXPERIMENTAL RESULTS}

In this section, we first verify that our proposed all-in-focus camera can be used for iris imaging DoF extension with a great capture volume. The DoF measurements of original telephoto lens and focus-tunable lens mounted after telephoto lens are compared across 1-5\,$m$ range. Following this analysis, we evaluate the Hamming distance of processed iris image during DoF extension at three distances. The spatiotemporal multiplexing imaging is demonstrated for auto-focusing among multiple participants within the extended capture volume. We conclude this section by describing the potential of our 3D light beam steering system in continuous iris imaging of a moving participant. 

\begin{figure}[!t]
\centering
\includegraphics[scale=0.32]{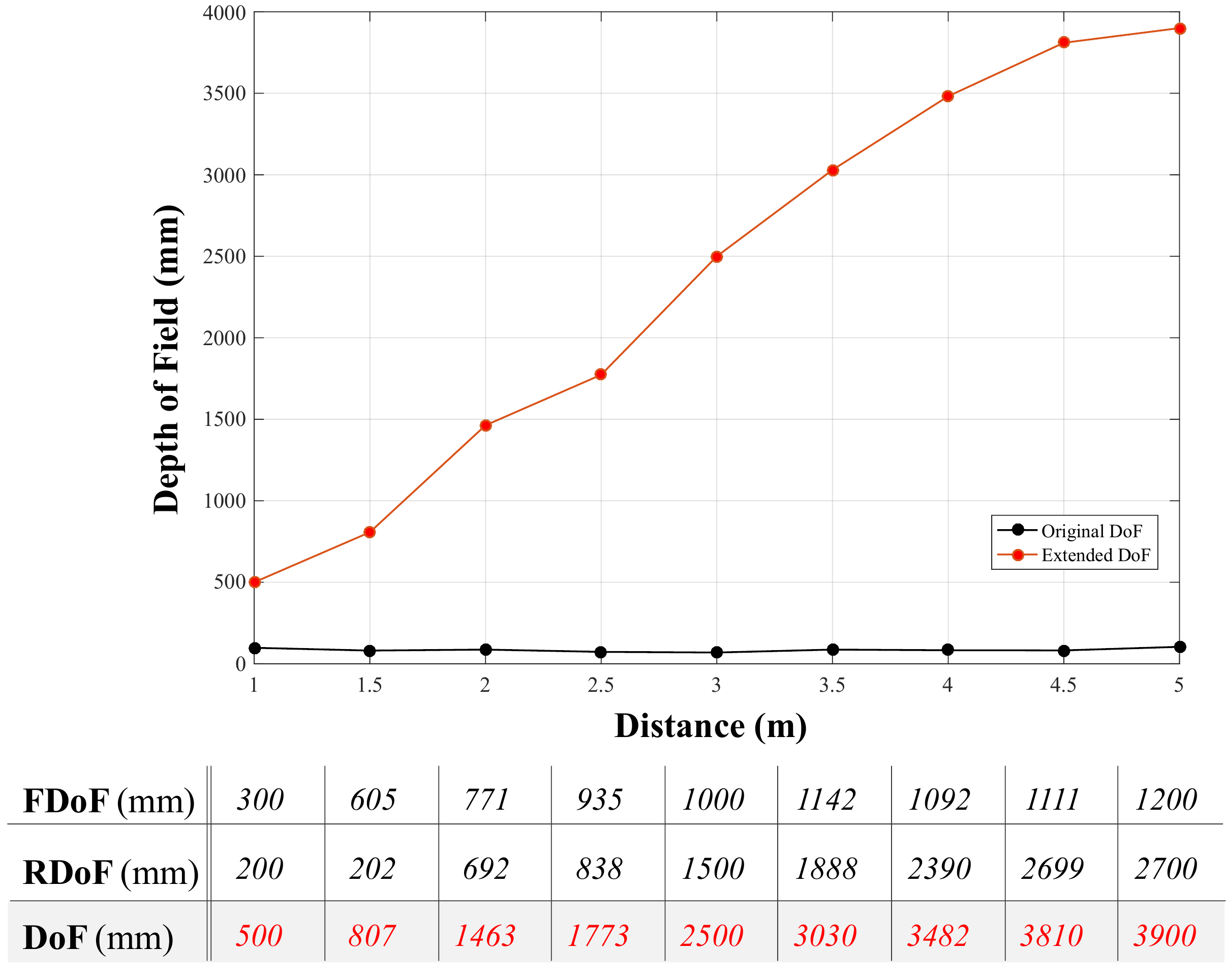}
\caption{\textbf{Evaluation of DoF extension capability by mounting the Optotune focus-tunable lens between telephoto zoom lens and image sensor.} The original telephoto zoom lens almost has a focal length less than 100\,$mm$ across the range. The total extended DoF is the sum of front DoF (FDoF) and rear DoF (RDoF). The DoF can be extended up to 3.9\,$m$ when the focus distance is 5\,$m$. The extended rear DoF is larger than the front DoF as expected. Each calculated DoF value shown in this image is the average of five repeated tests.}
\label{fig_dofextension}
\end{figure}

\begin{figure}[!ht]
\centering
\includegraphics[scale=0.52]{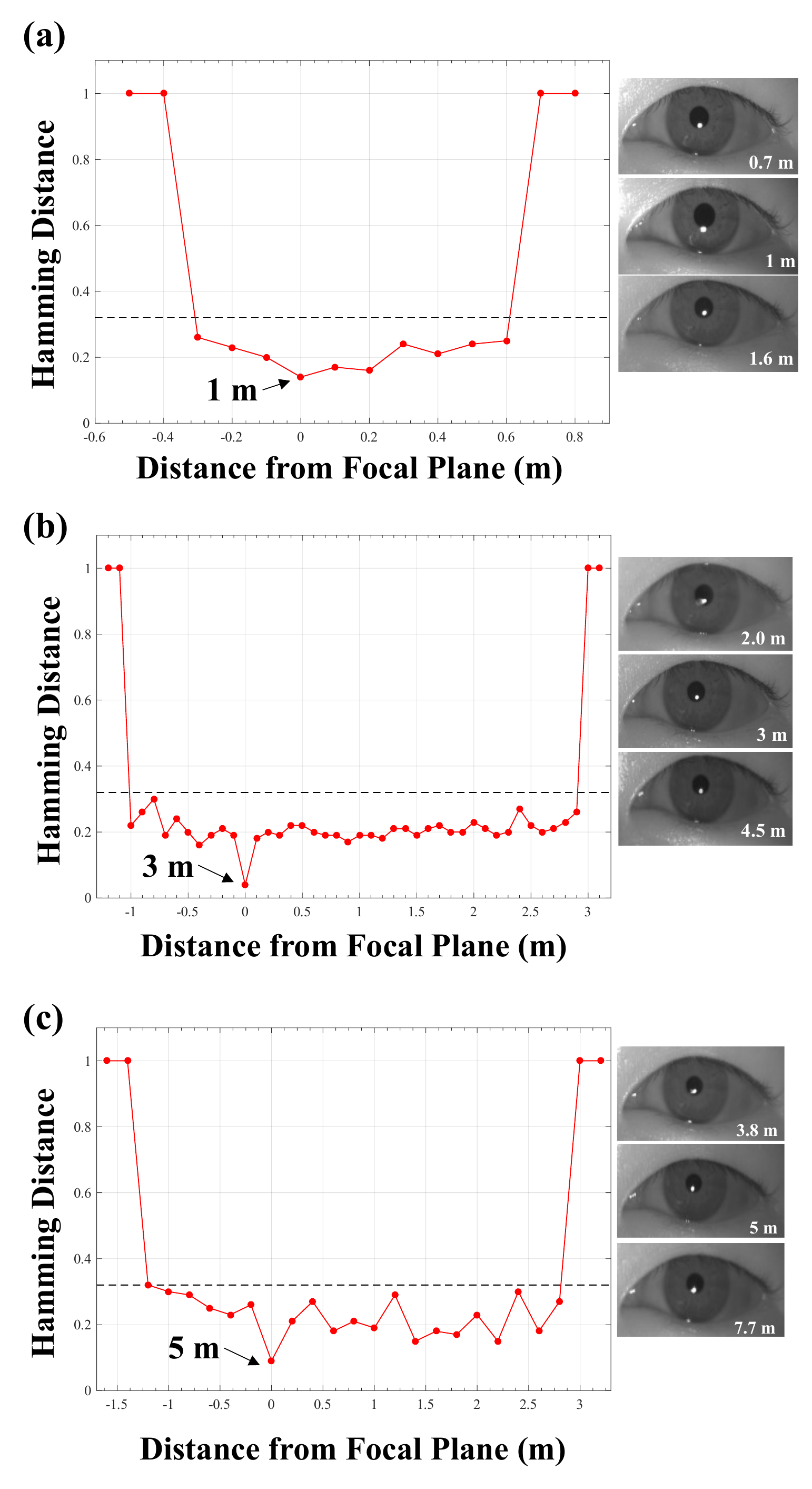}
\caption{\textbf{Hamming distance evaluation at various defocus distances during DoF extension.} Three DoF extension tests are conducted at different focus distances where (a) focal plane at 1\,$m$, (b) focal plane at 3\,$m$, and (c) focal plane at 5\,$m$. The image with an HD value below the dash line HD = 0.32 is considered as a match to the template image captured at the original focal plane. Both . The extended DoFs are approximately 0.9\,$m$ at 1\,$m$ distance, 2.5\,$m$ at 3\,$m$ distance and 3.9\,$m$ at 5\,$m$ distance respectively when we apply HD for evaluation. All images are evaluated without any enhancement processing.}
\label{fig_Hamming}
\end{figure}

\subsection{DoF Extension Measurement}
To evaluate the DoF extension capability, we conduct an experiment to calculate the actual DoF after focus-tunable lens is mounted compared with the original DoF of the telephoto lens. The image quality is measured according to the ISO standard in which the lateral resolution across the iris diameter should be at least 200 pixels. (In most cases the iris region resolution $\geq$150 pixels would be acceptable. Here we use a higher threshold for good recognition performance.) Every single image also has to pass the iris image quality test in which a self-developed algorithm in accordance with ISO standard is applied. The participant was asked to stand at 9 testing positions from 1 to 5\,$m$ with a 0.5\,$m$ interval to repeat the test 5 times. Starting from one designated position, we will control the focal plane of the telephoto zoom lens and the focus-tunable lens as the participant moves back and forth. Iris image quality check is conducted after all testing images are saved during post-processing. 

It is noted that the 200 pixel requirement mainly restricts the rear DoF extension because the iris image resolution decreases when liquid lens control current is adjusted from 0\,$mA$ to negative values. And the image quality requirement majorly regulates the front DoF extension. Because when we adjust the liquid lens control current from 0\,$mA$ towards larger positive value, the astigmatism is getting worse even the iris image is still focusable leading to blurred and distorted capture. The maximum DoF extension appears when the largest focal length 350\,$mm$ is used to capture the target at 5\,$m$ distance as shown in Figure \ref{fig_dofextension}. The extended DoF value 3900\,$mm$ (front DoF = 1200\,$m$ and rear DoF = 2700\,$mm$) is 37.5 times of the original DoF 104\,$mm$ from the telephoto lens. It takes about 80\,$ms$ to control the liquid lens focal plane moving through the whole extended DoF range until the higher order oscillations fully settled. If applying a low-pass filtered step signal, the settling time can be further reduced by 50\,$\%$ which is close to the time-consuming for 1 frame in real-time capture.

\subsection{Analysis of Hamming Distance}
Since an acceptable iris images has to be recognizable, the image quality evaluation of iris capture after DoF does not represent the actual extension capability from the perspective of recognition algorithm. In this study, we use the normalized Hamming distance (HD) which determines the similarity of two iris templates as the metric to evaluate the actual recognizable DoF range \cite{Daugman_Iris}. Following the general iris recognition framework, we preprocess the acquired iris images including segmentation, normalization, and feature extraction using our self-developed methods. After iris features are quantized into binary codes, the normalized HD is introduced as a quantitative measurement tool by bit-wise operation between two iris codes. Smaller HD values mean that the two iris image are from the same class with a greater probability. And if the HD value is higher indicating a lower likelihood of the two images belong to the same class. According to Daugman's extensive investigation \cite{Daugman_Statistic}, the HD value must be below 0.32 to statistically minimize the possibility of a false match between an enrolled iris template and a test iris template. 

\begin{figure}[!t]
\centering
\includegraphics[scale=0.4]{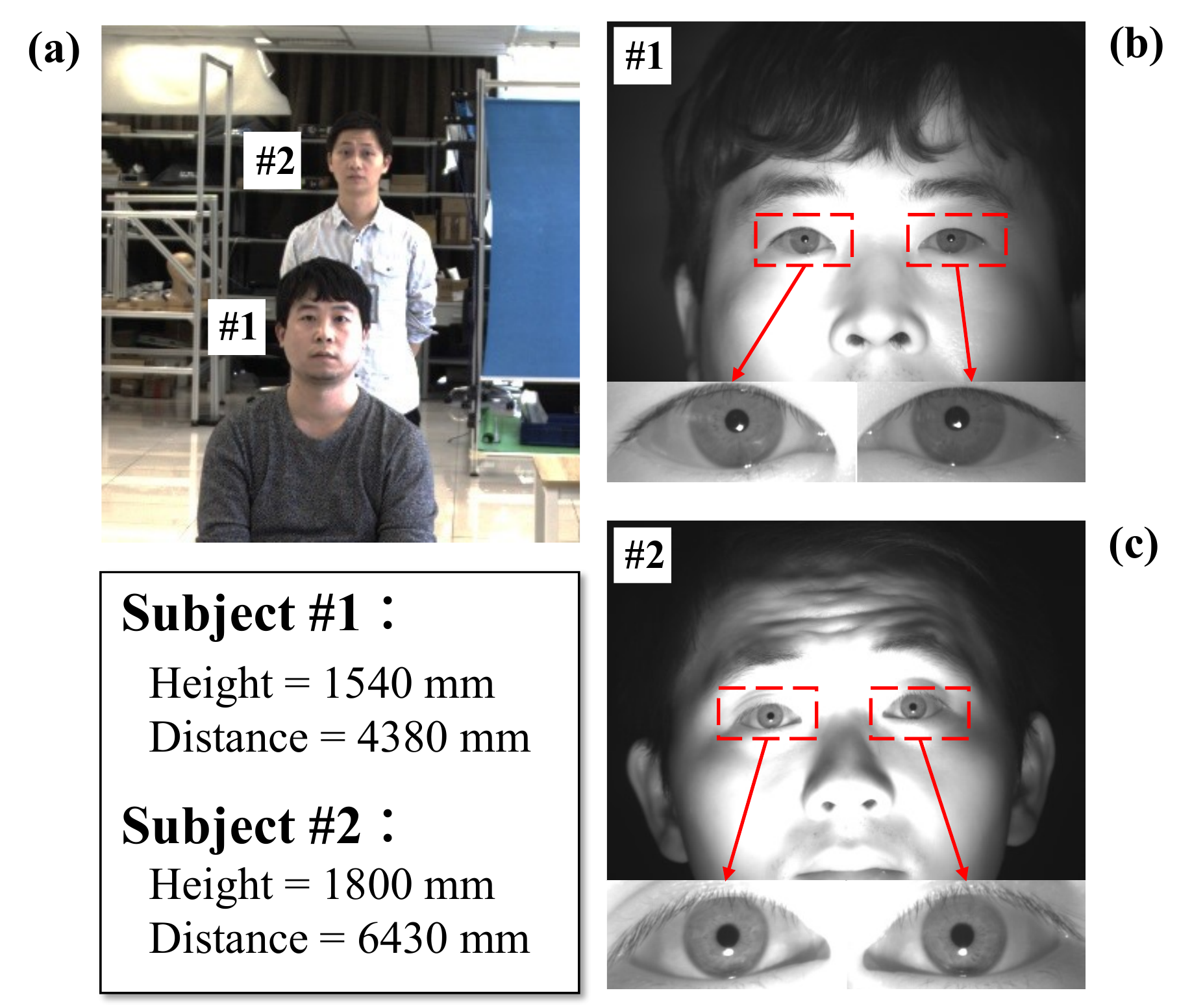}
\caption{\textbf{Demonstration of iris auto-refocusing between two participants for height and distance adaption.} High quality iris images are captured for two participants at two positions with different heights by jointly adjusting the focus-tunable lens and the 2D steering mirror (a). Subject \# 1 is sitting at 4380\,$mm$ with a height of 1540\,$mm$ (b) and subject \# 2 is standing at 6340\,$mm$ with a height of 1800\,$mm$ (c). Image is cropped for better demonstration which is not in accordance with the dimension scale.}
\label{fig_Refocusing}
\end{figure}

\begin{figure*}[!h]
\centering
\includegraphics[scale=0.46]{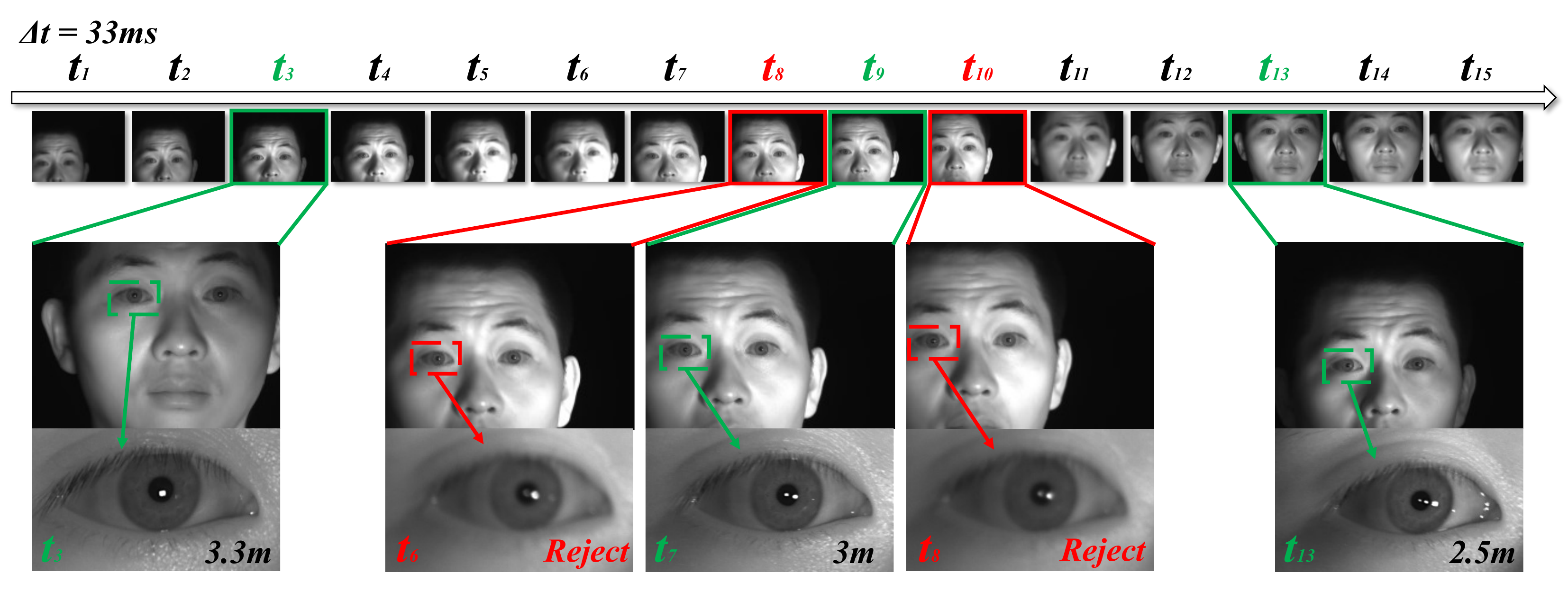}
\caption{\textbf{Demonstration to capture iris on the move using our proposed imaging system.} 15 sequential frames are acquired while the participant moves at a constant 1\,$m/s$ speed towards the imaging system. Three qualified iris images are captured at 3.3\,$m$, 3\,$m$ and 2.5\,$m$ as shown. The time interval between two captures is $\Delta$t = 33\,$ms$. The exposure time is set as 3\,$ms$ to sharply catch such a dynamic scene with proper illumination.}
\label{fig_IOM}
\end{figure*}

Here we conduct an experiment to analyze DoF extension capability at 1\,$m$, 3\,$m$ and 5\,$m$. The iris image captured at the best focal position as treated as the iris template enrollment and the iris images captured at the defocus positions ($\pm$0.1\,$m$ interval) are the test templates to compare with the enrollment. Figure \ref{fig_Hamming} shows the HD values when the test subject is standing at various defocus positions. It is noted that every HD value is calculated from the average of 5 repeated tests. The red point with a HD value below the 0.32 threshold means that iris template matches the enrolled iris template. Even with careful control, the image quality variation of each test does exist due to lighting, face pose and body movement. As expected, a larger DoF is derived compared with the DoF measurement based on image quality standard.

\subsection{Multi-Person Iris Auto-Refocusing}
By combining the focus-tunable lens and 2D steering mirror together, it is possible to capture the iris images of multiple participants at different locations with one camera. The fast response of our system enables the possibility of real-time multi-person iris imaging. We captured a scene of two people locating at different distance (4380\,$mm$ and 6430\,$mm$) from the imaging system with different heights (1540\,$mm$ and 1800\,$mm$) as shown in Figure \ref{fig_Refocusing}. As long as the distance and angle information fed into the control system, the steering mirror and liquid lens are simultaneously adjusted to direct the focused light rays of the target eyes on to imaging plane. Our system overcomes the obstacles such as height limit and position restriction in traditional long-range iris imaging system. It is experimentally found that iris image sequence capture is necessary in order to acquire at least one qualified testing sample. Besides the time consumption in hardware control, the efficiency of preprocessing methods including depth measurement, person detection and image quality evaluation also plays a major role in system performance. Further improvement could also focus on optimization of capture sequence strategy based on dynamic scene understanding.

\subsection{Applications in Iris Recognition}
In the same duration of time, larger capture volume allows more information collection of participants which increase the throughput of a biometric recognition system. This is also the research motivation of the first Iris on the Move (IOM) system back to 2006 \cite{Matey_IOM}. To make full use of the advantages of our 3D light beam steering iris imaging system, we also conduct an experiment to capture the iris image of a moving subject. We use the novel capture volume expanded camera with adaptively integrated lighting to replace the IOM design of lens with fixed focal distance and infrared lighting door. While a participant walks at a constant speed (around 1\,$m/s$) within the permissible DoF, our system actively captures a sequence of iris image frames in a continuous manner. Figure \ref{fig_IOM} shows that 3 focused frames are grabbed out of the 15 sequential frames during the walking period. If we look into the frames right before and after the focused frame $t_7$ at 3\,$m$, both $t_6$ and $t_8$ frames are blurry due to the unpredictable head movement while walking. The pivotal issue here is the imaging control algorithm for successful capture. The sensor exposure time needs to be an appropriated value. Long-exposure will introduce motion blur in a dynamic scene and short-exposure will transport insufficient photons to image sensor chip. In addition, we used a combination of fast focal sweep and auto-focusing control methods to follow the head motion trajectory of our target participant. The overview of this all-in-focus iris imaging system and demo videos including DoF extension, fast focal plane switch by manipulating focus-tunable lens, multi-person iris auto-refocusing recognition, and iris recognition of a moving participant are released on our lab website.

Despite the relatively rudimentary control algorithms, our results demonstrate the future of novel optical design with multiplexing imaging in iris recognition applications. Compared with the IOM system, our capture volume extended system provides a much higher throughput value towards unconstrained environments. It is believed that 3D light beam steering imaging method has potential applications in biometric recognition such as iris recognition on the move, continuous recognition, multi-person iris recognition, and omni-directional iris recognition. 

\section{CONCLUSION}

In this paper, we demonstrate an active iris capture system with the greatly extended DoF up to 3.9\,$m$ and the FoV up to $\pm$180\degree (H) x $\pm$60\degree (V). We develop this prototype all-in-focus iris camera by combining a two-axis beam steering mirror and the focus-tunable lens integrated with the inverted telephoto zoom lens. Our large aperture deformable optics provides high resolution iris image across a wide depth range without sacrificing photons transport. The fast response time of this electrically tunable lens enables real time refocusing by controlled focal sweeps. To capture iris images at various spatial positions in a temporal multiplexing way, we engineered a motorized 2D reflection mirror with precise and fast beam control which outperforms traditional iris imaging methods in FoV extension. Besides the great capture volume extension, our imaging system achieves high spatial, angular and temporal resolutions because of innovative hardware design. Experimental results show that the extended DoF is a factor of 37.5 compared with a conventional long focal lens when focus distance is 5\,$m$. And the Hamming distance analysis further confirms a larger DoF range does not sacrifice the quality of iris image for recognition. Our proposed system can dramatically increase the throughput of a biometric recognition system since it is verified in multi-person eye auto-refocusing and iris on the move applications.

Our research concludes that this iris imaging volume extension camera has a great potential to inspire novel biometric applications such as continuous recognition, omni-directional recognition, and active recognition. We expect that the performance of our iris imaging system can be further improved by introducing adaptive illumination unit and replacing steering module with a MEMS mirror for agile capture control. In the following study, we have the plan to develop precise synchronous control method and efficient scene understanding algorithms such as object detection, depth measurement, pose estimation, and semantic segmentation. It is also necessary to collect 
sufficient iris data in different scenarios to verify the performance of our all-in-focus system statistically. The multi-person iris recognition on the move and iris recognition in surveillance scene will be other interesting topics to explore using spatiotemporal multiplexing imaging method.

\section*{Acknowledgements}
This work was supported by the National Key Research and Development Project (Grant No.2017YFB0801900), Tianjin Key Research and Development Project (Grant No.17YFCZZC00200) and Science and Technology Cooperation Project with Academy and University of Sichuan Province (Grant No.18SYXHZ0015) .

{\small
\bibliographystyle{ieee}
\bibliography{reference}
}

\end{document}